\documentclass[12pt]{article}
\usepackage[]{graphicx}
\usepackage[]{color}
\makeatletter
\def\maxwidth{ %
	\ifdim\Gin@nat@width>\linewidth
	\linewidth
	\else
	\Gin@nat@width
	\fi
}
\makeatother

\definecolor{fgcolor}{rgb}{0.345, 0.345, 0.345}

\usepackage{framed}
\makeatletter
{\par\unskip\endMakeFramed%
	\at@end@of@kframe}
\makeatother

\definecolor{shadecolor}{rgb}{.97, .97, .97}
\definecolor{messagecolor}{rgb}{0, 0, 0}
\definecolor{warningcolor}{rgb}{1, 0, 1}
\definecolor{errorcolor}{rgb}{1, 0, 0}
\usepackage{alltt}%\documentclass[nojss]{jss}

\usepackage[latin1]{inputenc}
\usepackage{amsfonts}
\usepackage{amssymb}
\usepackage{amsmath}
\usepackage{amsthm}
\usepackage{graphicx,psfrag,epsf}
\usepackage{enumerate}
\usepackage{bbm}
\usepackage{float}
\usepackage[authoryear,sort&compress]{natbib}
\newcommand{\blind}{0}

\IfFileExists{upquote.sty}{\usepackage{upquote}}{}
\begin{document}
	%\SweaveOpts{concordance=TRUE}
	\def\spacingset#1{\renewcommand{\baselinestretch}%
		{#1}\small\normalsize} \spacingset{1}	
	
	\large
	
	\newcommand{\al}{\mbox{$\alpha$}}
	\newcommand{\be}{\mbox{$\beta$}}
	\newcommand{\ep}{\mbox{$\epsilon$}}
	\newcommand{\gam}{\mbox{$\gamma$}}
	\newcommand{\sig}{\mbox{$\sigma$}}
	
	\DeclareRobustCommand{\FIN}{%
		\ifmmode % if math mode, assume display: omit penalty etc.
		\else \leavevmode\unskip\penalty9999 \hbox{}\nobreak\hfill
		\fi
		$\bullet$ \vspace{5mm}}
	
	\newcommand{\calA}{\mbox{${\cal A}$}}
	\newcommand{\calB}{\mbox{${\cal B}$}}
	\newcommand{\calC}{\mbox{${\cal C}$}}
	
	\newcommand{\muas}{\mbox{$\mu$-a.s.}}
	\newcommand{\Nat}{\mbox{$\mathbb{N}$}}
	\newcommand{\Rea}{\mbox{$\mathbb{R}$}}
	\newcommand{\Prob}{\mbox{$\mathbf{P}$}}
	\newcommand{\ProbQ}{\mbox{$\mathbf{Q}$}}
	
	\newcommand{\nin}{\mbox{$n \in \mathbb{N}$}}
	\newcommand{\suc}{\mbox{$\{X_{n}\}$}}
	\newcommand{\sucP}{\mbox{$\mathbb{P}_{n}\}$}}
	
	\newcommand{\conv}{\rightarrow}
	\newcommand{\convn}{\rightarrow_{n\rightarrow \infty}}
	\newcommand{\convp}{\rightarrow_{\mbox{c.p.}}}
	\newcommand{\convs}{\rightarrow_{\mbox{a.s.}}}
	\newcommand{\convw}{\rightarrow_w}
	\newcommand{\convd}{\stackrel{\cal D}{\rightarrow}}
	\newcommand{\R}{\mathbb{R}}
	\newcommand{\Rn}{\mathbb{R}^n}
	\newcommand{\PR}{\mathbb{P}}
	\newcommand{\Rd}{{\displaystyle\mathbb{R}^2}}
	\newcommand{\Rb}{\mathbb{\overline{R}}}
	\newcommand{\Rbd}{{\displaystyle \mathbb{\overline{R}}^2}}
	\newcommand{\Rbn}{{\displaystyle \mathbb{\overline{R}}^n}}
	\newcommand{\I}{\mathbb{I}}
	\newcommand{\Sv}{\mathbb{S}}
	\newcommand{\Id}{{\displaystyle\mathbb{I}^2}}
	\newcommand{\In}{{\displaystyle\mathbb{I}^n}}
	\newcommand{\Z}{\mathbb{Z}}
	\newcommand{\N}{\mathbb{N}}
	\newcommand{\Hbb}{\mathbb{H}}
	\newcommand{\lp}{\left(}
	\newcommand{\rp}{\right)}
	\newcommand{\lc}{\left[}
	\newcommand{\rc}{\right]}
	\newcommand{\lb}{\left\{}
	\newcommand{\rb}{\right\}}
	\newcommand{\lf}{\left.}
	\newcommand{\ri}{\right.}
	\newcommand{\id}{\stackrel{d}{=}}
	\newcommand{\prob}[1]{\PR\lp#1\rp}
	\newcommand{\esp}[2]{\mathbf{E}_#1\lc#2\rc}
	\newcommand{\espe}[1]{\mathbf{E}\lc#1\rc}
	\newcommand{\cb}{C\!\!\!\!/\!\!\!/\,}
	\newcommand{\notprec}{\not\prec}
	\providecommand{\abs}[1]{\left|#1\right|}
	\newcommand{\casos}[4]{\lb\begin{array}{ll}#1,&#2\\#3,&#4\end{array}\ri}
	\newcommand{\caso}[2]{\lb\begin{array}{l}#1\\#2\end{array}\ri}
	\newcommand{\sign}{\textrm{sign}}
	\newcommand{\corr}[2]{\textrm{Corr}\lp#1,#2\rp}
	\newcommand{\cov}[2]{\textrm{Cov}\lp#1,#2\rp}
	\newcommand{\Cov}[1]{\textrm{Cov}\lp#1\rp}
	\newcommand{\var}[1]{\textrm{Var}\lp#1\rp}
	\newcommand{\lcm}{\text{lcm}}
	\setlength{\parindent}{0in}
	
	% Comands with arguments
	\newcommand{\lrp}[1]{\left(#1\right)}
	\newcommand{\lrc}[1]{\left[#1\right]}
	\newcommand{\lrb}[1]{\left\{#1\right\}}
	\newcommand{\E}[1]{\mathbf{E}\lc #1\rc}
	\newcommand{\V}[1]{\mathbb{V}\mathrm{ar}\lc #1\rc}
	\newcommand{\Es}[2]{\mathbf{E}_{#2}\lc #1\rc}
	\newcommand{\Vs}[2]{\mathbb{V}\mathrm{ar}_{#2}\lc #1\rc}
	\newcommand{\mse}[1]{\mathrm{MSE}\lrc{#1}}
	\newcommand{\mise}[1]{\mathrm{MISE}\lrc{#1}}
	\newcommand{\amise}[1]{\mathrm{AMISE}\lrc{#1}}
	\newcommand{\pf}[2]{\frac{\partial #1}{\partial #2}}
	\newcommand{\pftwo}[2]{\frac{\partial^2 #1}{\partial #2^2}}
	\newcommand{\pfmix}[3]{\frac{\partial^2 #1}{\partial #2\partial #3}}
	\newcommand{\norm}[1]{\left|\left| #1\right|\right|}
	\newcommand{\tr}[1]{\text{tr}\left[#1\right]}
	\newcommand{\inprod}[2]{\langle#1,#2\rangle}
	\newcommand{\vlinel}[1]{\multicolumn{1}{|c}{#1}}
	\newcommand{\vliner}[1]{\multicolumn{1}{c|}{#1}}
	
	\newcommand{\bb}[1]{\mathbb{#1}}
	\newcommand{\mcal}[1]{\mathcal{#1}}
	\newcommand{\mat}[1]{\mathbf{#1}}
	\newcommand{\ind}[1]{\mathbbm{1}_{\lrb{#1}}}

	\newtheorem {Prop}{Proposition} [section]
	\newtheorem {Lemm}[Prop] {Lemma}
	\newtheorem {Tabla}[Prop] {Table}
	\newtheorem {Theo}[Prop]{Theorem}
	\newtheorem {Coro}[Prop] {Corollary}
	\newtheorem {Nota}{Remark}[Prop]
	\newtheorem {Ejem}[Prop] {Example}
	\newtheorem {Defi}[Prop]{Definition}
	\newtheorem {Figu}[Prop]{Figure}
	\if0\blind
	{
		\title{\bf Predicting seasonal influenza transmission using Regression Models with Temporal  Dependence}
\author{	Manuel Oviedo de la Fuente\textsuperscript{1,2*}, Manuel Febrero--Bande\textsuperscript{1}, \\
María Pilar Mu\~{n}oz\textsuperscript{3} and Àngela Dom\'{i}nguez\textsuperscript{4,5}
}

\maketitle

\footnotetext[1]{  Dept. of Statistics and Op. Res. Universidade de Santiago de Compostela, Spain.}
\footnotetext[2]{	 Instituto Tecnológico de Matemática Industrial (ITMATI), Spain.}
\footnotetext[3]{		Dept. of Statistics and Op. Res.,  Universitat Politècnica de Catalunya, Spain.}
\footnotetext[4]{		Dept. of Medicine. Universitat de Barcelona, Spain.}
\footnotetext[5]{	CIBER en Epidemiolog\'{i}a y Salud  P\'{u}blica (CIBERESP), Spain.
	
\hspace{.0cm} *Corresponding author. e-mail: manuel.oviedo@usc.es.}

}
	
\begin{abstract}	
	In this manuscript, we use meteorological information in Galicia (Spain) to propose a novel approach to predict the incidence of influenza. %	 we propose a novel approach to predict the incidence of influenza  using meteorological information in Galicia (Spain).	 
	  Our approach extends the GLS methods in the multivariate framework  to functional regression models with dependent errors.   %A simulation study shows that the GLS estimators improve the classical linear approach with better estimations of the parameters associated with the regression model and extremely good results from the predictive point of view.	  
	  A simulation study shows that the GLS estimators render better estimations of the parameters associated with the regression model and obtain extremely good results from the predictive point of view. Thus they improve the classical linear approach. It proposes an iterative version of the GLS estimator (called  iGLS) that can help to model complicated dependence structures, uses the distance correlation measure  $\mathcal{R}$ to select relevant information to  predict influenza rate and applies the GLS procedure to the prediction of the influenza rate using readily available functional variables.  These kinds of models are extremely useful to health managers in allocating resources in advance for an epidemic outbreak.
	  %	  These kinds of models are particularly useful to health managers who may use them to allocate resources for an epidemic outbreak in advance.% 	  In particular, the nice interpretation of the model shows that influenza may increase  due to a cold wave with average daily temperatures around $5$ C.

\end{abstract}


	
%	62J12  	Generalized linear models
%	62H12  	Estimation
% 	62M10  	Time series, auto-correlation, regression, etc. [See also 91B84]
%	62M20  	Prediction [See also 60G25}; filtering \SeeAlso{60G35, 93E10, 93E11]}
% 	62P10  	Applications to biology and medical sciences

%\begin{keyword}
	%Climatological variables; Dependent data; Functional data analysis; Influenza;	correlated errors
	%\end{keyword}
	
\spacingset{1.45}

\section{Introduction}

Influenza is an infectious disease with person-to-person transmission that characteristically occurs as an epidemic affecting the whole population, see for instance, \cite{Watson2012}. The influenza virus has been categorized into types A, B and C. However influenza C is a mild disease without seasonality and is therefore not considered in  influenza epidemics. One remarkable feature of the influenza A and B viruses is the frequency of changes in antigenicity. Alterations in the antigenic structure of the virus leads to infection by variants to which the population has little or no resistance. 

The epidemiology of inter-pandemic influenza (also named seasonal influenza) is characterized in temperate zones by epidemics of variable size that occur during the colder winter months (November to April in the Northern Hemisphere and May to September in the Southern Hemisphere), each of which typically lasts 8-10 weeks (\cite{VanTam2013}). In a study on influenza activity throughout eight seasons (1999--2007), the average length of epidemics in 23 European countries was 15.6 weeks (median 15 weeks; range 12-19 weeks), see \cite{paget2007influenza}.% However, cases and outbreaks in institutional settings may occur year round (\cite{VanTam2013}). 

The reasons for the seasonal presentation of epidemics are not entirely clear but they might result from more favorable environmental conditions for virus survival (\cite{Schaffer1976}). Various theories including improved virus survival in low temperatures, low humidity and low levels of ultraviolet radiation (\cite{VanTam2013}) have been advanced to explain this pattern in temperate zones. The typical incubation period for influenza is 1-4 days (average: 2 days). 

Surveillance systems require accurate indicators that detect possible epidemics in advance. The epidemic of influenza is one of the problems of most concern to public health professionals across the world, due to its high levels of mortality and morbidity. Influenza is highly contagious and causes more morbidity than any other vaccine-preventable illness (\cite{monto2002zanamivir}). So accurate estimates of the incidence of influenza are essential, for both public health services and citizens,  to provide advance warning of epidemics and allow preventive measures to reduce contagion.
%Statistical methods of forecasting the incidence of influenza in particular and contagious diseases in general has changed over time. One of the first studies on time series, by \cite{choi1981evaluation}, fit an ARIMA model for estimating pneumonia and influenza mortality to determine the number of deaths they caused. \cite{Dushoff2006} investigated how cold temperatures contribute to excess seasonal mortality through a regression model. An alternative model of  monitoring infectious diseases was proposed by \cite{hohle2008count}, and from a Bayesian approach in Conesa et al. (2015).  Those studies consist of applying count data charts to monitor time series, such as  automated monitoring of influenza surveillance data.\\

Statistical methods to forecast the incidence of influenza in particular, and contagious diseases in general, have changed over time. In one of the first studies on time series, \cite{choi1981evaluation} employed an ARIMA model to estimate pneumonia and influenza mortality. \cite{Dushoff2006} used a regression model to investigate how cold temperatures contribute to excess seasonal mortality. \cite{hohle2008count} proposed an alternative model to monitor infectious diseases that consisted in applying count data charts to monitor time series.  From a Bayesian framework,  \cite{Conesa2015} proposed automated monitoring of influenza surveillance data that made it possible to take the geographical component into account in statistical models in addition to temporal evolution. Contributions to this methodology are growing steadily through disease mapping. The studies by \cite{ugarte2010spatio} and \cite{paul2011predictive} are recent examples of this. Their common denominator  is that they apply different statistical methodologies to multivariate time series (hierarchical Bayesian space--time, mixed models, P--splines and conditional autoregressive models (CAR), among others) of infectious disease counts, collected in different geographic areas, using multivariate or longitudinal data. 

Functional data analysis (FDA) has grown in popularity over recent years alongside the increasing availability of continuous measurements in different contexts like  Biomedicine (\cite{Cuevas2004}), Spectrometry (\cite{Ferraty2006}), Biology (\cite{Chiou2003}) and Medicine (\cite{Sorensen2013}), to mention only a few. This study extends the regression models for independent functional data to the case where the curves presents either spatial or temporal  dependencies.

% Fundir párrafo
Our goal is to estimate the rate of influenza epidemics, using the information readily available from public sources possibly that include functional variables, by adapting or extending the GLS techniques from a multivariate framework to this new framework. %This studies introduces a functional model which includes both spatial and temporal components.
So, our particular aim is to estimate the spatial or temporal dependence  components of influenza, using regression models, and predict the rate of incidence of influenza for a horizon of 14 days (2 weeks). We initially model influenza using a traditional linear approach (with independent errors) and later extend these ideas to the functional case (with dependent errors). 
%Fin  Fundir párrafo

The article is structured as follows. Section~\ref{methods} presents the GLS approach for functional regression models. The estimation of the different parameters (for the regression function or the dependence) is usually done using maximum likelihood although, as an alternative, we introduce an iterative GLS (iGLS) procedure that provides similar results. The latter could be interesting when the structure of the dependence is complicated. The practical performances of the  GLS and iGLS procedures  are compared, by means of a simulation study (Section~\ref{simulation}), to the case where the dependence is not considered.  Section~\ref{example} applies these models to the prediction of the influenza rate in a region of Spain. Finally, Section~\ref{conclusion} discusses the results obtained.

%------------------------------------------------------------------
\section{Methodology}\label{methods}

The functional regression model (FRM) is one of the most studied topics in FDA over the last few years. A regression model is said to be ``functional'' if any of the variates involved (the predictors or the response) has a functional nature, i.e. it is a measure observed along a continuous interval. Cases with a scalar response and functional predictors have particularly attracted a lot of attention. For example, \cite{Sorensen2013} give a basic  introduction  for the analysis of functional data applied in datasets from medical science.%Sorensen et. al 2013 is a review of some of the techniques developed for functional data applyied in datasets from medical science. from biomedicine dataset

The functional regression model with scalar response (FRM) is stated as follows:\newline
Let $\mcal{X},Y$ two random variates taking values in $\mcal{E}\times\R$ where $\mcal{E}$ is a functional space (semi-metric, normed or Hilbert). The relationship between the two variates can be expressed as follows:
\begin{equation}
	Y=m\lrp{\mcal{X}}+\epsilon=\E{Y|\mcal{X}}+\epsilon
	\label{frm}
\end{equation} 
where $\epsilon$ is a real random variable verifying $\E{\epsilon|\mcal{X}}=0$. Depending on the nature of the functional space $\mcal{E}$ and on the regression operator $m$, we can classify the different types of FRM:

\begin{itemize}
	\item \textbf{Multivariate Linear Model}: $\mcal{E}=\R^p$ and $m$ is the linear operator in the space, i.e. $\E{Y|\mcal{X}}=\mcal{X}\beta$ with $\beta\in\R^p$.
	\item \textbf{Functional Linear Model}:  $\mcal{E}=\mcal{L}^2(T)$ is the Hilbert space of square integrable functions over $T$ and $m$ is a linear operator in the space, i.e. $m\lrp{\mcal{X}}=\inprod{\mcal{X}}{\beta}$ with $\beta\in\mcal{L}^2(T)$. This model has been treated extensively in the literature mainly devoted to the optimal way of representing the linear operator through the representation of $\mcal{X}$ and $\beta$ on a basis of $\mcal{L}^2(T)$. % A ?good? basis should be parsimonious in the sense that a large set of possible response functions can be approximated well by only few terms of the	generalized Fourier expansion employed.
	
	Depending on the latter, the references can be classified into two main categories: 
	\begin{itemize} 
		
		\item Fixed basis.  The most commonly used basis in this context are the Fourier (see \cite{Ramsay2005}), the B-spline (see \cite{Cardot2003}) and the Wavelet (see for instance \cite{Antoniadis2003}).
		
		% Several approach provided an appropriate choice of basis to estimate  the unknown  function  $m$. In  \cite{Ramsay2005}  uses  Fourier coefficients from the data (orthogonal and properly for periodic data),  \cite{Cardot2003} uses spline estimators for the functional linear model (properly for non periodic data) and \cite{Antoniadis2003} uses the Wavelet basis  which allow a parsimonious expansion for a wide variety of functions (properly for inhomogeneous case). 
		
		%	\item	Fixed basis. The basis expansions has been studied and improved by several authors including Fourier basis in \cite{Ramsay2005}, B-spline basis in \cite{Cardot2003} and  the Wavelet basis  in \cite{Antoniadis2003}, to cite the most used.		
		
		\item Data-driven basis. Two main basis computed from the data are used in the literature: the most parsimonious one is given by the functional principal components (see \cite{Horvath2012} and \cite{Cardot1999}) and the one that maximizes the covariance among the response and the functional predictor uses the functional partial least square components (PLS) (see for instance \cite{Cardot2007a} and \cite{Preda2005}).
		
		%Data-driven basis. Two main procedures are use in literature: (1) the functional principal components (FPC) represents functional data in the most parsimonious way, in the sense that when using a fixed number of basis functions, see \cite{Horvath2012} and  \cite{Cardot1999}, and  (2) the functional partial least squares (PLS) which maximize the variance between the response and functional predictor, see \cite{Cardot2007a} and \cite{Preda2005}.
	\end{itemize}
	
	Note that, due to the representation employed, the FRM is always an approximated model and its goodness typically relies on the  properties of the chosen basis and its suitability to the data at hand. 
	
	\item \textbf{Functional Non Linear Model}: $\mcal{E}$ is (at least) a semi-metric space and $m$ is a continuous operator i.e. $\mathop{lim}_{\mcal{X}'\rightarrow\mcal{X}}m(\mcal{X}')=m(\mcal{X})$. For a complete review of this model see \cite{Ferraty2006} and the references therein.
	\item \textbf{Extensions of the above models}: The above models could be extended in several ways, usually considering more than one predictive variate. This could lead to semi-linear models (\cite{Aneiros2006, Aneiros2008}), additive models (\cite{MullerYao2008}, \cite{Ferraty2009}, \cite{Febrero2013}), single index models (\cite{Chen2011}, \cite{Goia2012}) or projection pursuit models (\cite{Ferraty2013}).
\end{itemize}

Many of the above-mentioned authors consider that $\epsilon=\left(\epsilon_1,\ldots,\epsilon_{n,s}\right)'$ is an homoskedastic independent error vector, i.e. $\E{\epsilon}=0, \V{\epsilon}=\sigma^2$ and $\Cov{\epsilon_i,\epsilon_j}=0, i\ne j$. This assumption is made to obtain simple diagnostics or confidence intervals for the response but it could be too restrictive in functional regression models and difficult to check or fulfill in practice. Some papers consider dependence in the functional variate. See, for example, \cite{Delicado2010}, \cite{Giraldo2011} and \cite{Menafoglio2013} for contributions devoted to spatial dependence with functional data or \cite{Battey2013}, \cite{Besse2000}, \cite{Damon2005} and \cite{Hormann2010} for time dependence. In both cases, the functional nature of the variate complicates the predictive ability of the model. The aim of this paper is to extend the GLS approach (see for example \cite{Kariya2004}) to the functional context as the simplest way of incorporating temporal or spatial dependence in the regression models. In fact, the GLS approach can handle a wide range of regression models with dependence in a simple way: equi-correlation models, random effects, time and spatial dependence, and so on.

%---------------------------
%-----------------
\subsection{Functional Generalized Least Squares Regression }
The functional generalized least squares regression (FGLS) model between two centered variables ($\E{y}=0$, $\E{\mcal{X}}=0$) states that
\begin{equation}
	y=\inprod{\mcal{X}}{\beta}+\epsilon = \int_{T}\mcal{X}(t)\beta(t)dt+\epsilon
	\label{fgls}
\end{equation}
where $\beta\in\mcal{L}_2(T)$ and $\epsilon$ is now a random vector with mean 0 and covariance matrix $\Omega=\E{\epsilon\epsilon^\prime}$.  
This model includes, as its special cases, many others models, all of them based on $\Omega=\Omega(\theta)=\sigma^2\Sigma(\theta)$, where $\theta$ is the parameter associated with the dependence structure of $\Omega$. Some classical examples are presented in the following models:

\begin{enumerate}[(a)]
	\item Equi-correlated model: $\V{\epsilon_i}=\sigma^2$ and $\cov{\epsilon_i}{\epsilon_j}=\sigma^2\theta, i\ne j, \theta \in \lrp{-1,1}$ 
	%$$
	%\Omega=\sigma^2\lrp{\begin{array}{cccc} 1 & \theta &\cdots& \theta \\ \theta & 1 & & \vdots \\ \vdots & & \ddots & \theta \\ \theta & \cdots & \theta & 1\end{array}}
	%$$
	\item Heteroskedastic block model: 
	%$$
	%\Omega=\lrp{\begin{array}{cccc} \sigma^2_1I_{n_1} & 0 & \cdots & 0 \\ 0 & \sigma^2_2I_{n_2} & 0 & \vdots \\ \vdots & & \ddots & 0 \\ 0 & \cdots & 0 & \sigma^2_pI_{n_p}\end{array}}
	%$$
	$
	\Omega=\mathop{diag}\lrp{\sigma^2_1\mat{I}_{n_1} | \sigma^2_2\mat{I}_{n_2} | \cdots | \sigma^2_p\mat{I}_{n_p}}
	$
	with $n_1+n_2+\cdots+n_p=n$
	\item AR(1) model: $\epsilon_i=\theta\epsilon_{i-1}+\varepsilon_i$ with $\abs{\theta}<1$, $\E{\varepsilon_i}=0$, $\V{\varepsilon_i}=\tau^2$ and $\cov{\varepsilon_i}{\varepsilon_j}=0, i\ne j$
	%$$
	%\Omega=\frac{\tau^2}{1-\theta^2}\lrp{\begin{array}{cccc} 1 & \theta & \cdots & \theta^{n-1} \\ \theta & 1 & \theta & \theta^{n-2} \\ \ddots & \ddots & \ddots & \vdots \\  \theta^{n-1} & \theta^{n-2} & \cdots & 1 \end{array}}
	%$$
	$$
	\Omega=\frac{\tau^2}{1-\theta^2}\lrp{\theta^{\abs{i-j}}}_{i,j=1}^n
	$$
	The variance structure is also known for every ARMA($p$,$q$) model.
	\item Spatial correlation model: 
	$$
	\Omega=\sigma^2\lrp{\rho\lrp{d(s_i,s_j)}}
	$$
	where $s_i$,$s_j$ are, respectively, the locations for $i,j$; and $\rho$ is the spatial correlation function.
\end{enumerate}

\subsection{Estimation of FGLS}\label{estimation}
The classical theory of \cite{Kariya2004} can be extended to the functional case by adapting the GLS criterion accordingly, i.e. 
$$\mathop{GLS}(\beta,\theta)=\lrp{y-\inprod{\mcal{X}}{\beta}}^\prime\Sigma(\theta)^{-1}\lrp{y-\inprod{\mcal{X}}{\beta}}$$

Given the sample $\lrb{(\mcal{X}_1,y_1),\ldots, (\mcal{X}_{n,s},y_{n,s})}$, we can approximate $\mcal{X}_i$ and $\beta$ using a finite sum of the basis elements: 
\[
%\mcal{X}_i(t)=\sum_{k}^{\infty}c_{ik}\psi_{k}(t)\approx\sum_{k}^{K_x}c_{ik}\psi_{k}(t), \; %\beta(t)=\sum_{k}^{\infty}b_{k}\varphi_{k}(t)\approx\sum_{k}^{K_\beta}b_{k}\varphi_{k}(t) 
\mcal{X}_i(t)\approx\sum_{k}^{K_x}c_{ik}\psi_{k}(t), \; \beta(t)\approx\sum_{k}^{K_\beta}b_{k}\varphi_{k}(t) 
\]
The preceding equations can be expressed as matrix notation using the evaluation in a grid of the length M $\{a=t_1<\cdots<t_M=b\}$ as  
$$\mat{X}=\mat{C}\Psi, \;\; \mat{B}=\mat{b}^\prime\varphi$$,  
where $\mat{X}$ is the matrix $n\times M$ with the evaluations of the curves in the grid, $\mat{C}$ is the matrix $n\times K_x$ with the coefficients of the representation in the basis and $\Psi$ is the matrix $K_x\times M$ with the evaluations of the basis elements on the grid. Similarly, $\mat{B}$ is the matrix ($1\times M$) with the evaluation of the $\beta$ parameter on the grid, $\varphi$ is the matrix ($K_\beta\times M$) with the evaluations of the basis $\{\varphi_j\}$ and $\mat{b}$  on the grid, is the vector of the coefficients of $\beta$ in the basis.

With this notation, the terms $\{\inprod{\mcal{X}_i}{\beta}\}_{i=1}^n$ can be approximated by $\mat{C}\Psi\varphi^\prime\mat{b}=\mat{Z}\mat{b}$ which, in essence, is a reformulation of a classical multivariate linear model that approximates the functional model. Here, the matrix $\mat{Z}$ takes into account all the approximation steps done with the information available: the chosen basis for $\mcal{X}$ and $\beta$ with the selected components: $K_x$ and $K_\beta$.

Once a certain approximation is selected, supposing that $\theta$ is known, we can define  $\mat{W}=\Sigma(\theta)^{-1}$, and use the classical theory for multivariate GLS to obtain the BLUE of $\mat{b}$ through:
\[
\mat{b}_\Sigma=\mat{(Z^\prime WZ)^{-1}Z^\prime W}y,\]
where $\mat{b}_\Sigma$ has covariance
\[
\Cov{\mat{b}_\Sigma}=\sigma^2\lrp{\mat{Z^\prime WZ}}^{-1}
\]

Finally, the fitted values are obtained by:
\[ \hat{y}=\mat{Z(Z^\prime WZ)^{-1}Z^\prime W}y=\mat{H}_\Sigma y\]
where $\mat{H}$ is the hat matrix.

Once the model is estimated, we can compute the prediction in a new point $\lrb{\mcal{X}_0}$ using the model chosen for $\Sigma$. Being $\Delta'=\cov{\epsilon}{\epsilon_0}$ and $\Sigma_0=\Cov{\epsilon_0\epsilon_0^\prime}$, we can obtain the equations for prediction:
\[
\hat{y_0}=\inprod{\mcal{X}_0}{\hat{\beta}}+\Delta\Sigma^{-1}\lrp{y-\inprod{\mcal{X}}{\hat{\beta}}}
\]
\[
\V{\hat{y}_0}=\sigma^2\lrp{\Sigma_0-\Delta\Sigma^{-1}\Delta^\prime}
\]

%\subsubsection{Estimation of $\theta$}

The GLS criterion can be employed to jointly estimate all the parameters associated to the model and can be expressed as: 
$$\mathop{min}_{K_x, K_\beta,\mat{b},\theta} \mathop{GLS}=\mathop{min}_{K_x, K_\beta, \mat{b},\theta}\lrp{y-\mat{Z}\mat{b}}^\prime\Sigma(\theta)^{-1}\lrp{y-\mat{Z}\mat{b}},$$
where the parameters $K_x$ and $K_\beta$ related to the basis for $\mcal{X}$ and $\beta$ are typically chosen \emph{a priori} taking into account, for instance, the quality of the data and its representation on the discretization grid or other considerations related to the data-generating process (smoothness, physical restrictions, interpretability,...). The direct minimization of GLS usually cannot be affordable even though we only consider the parameters  $\mat{b}$ and $\theta$. The generalized cross-validation (GCV) criterion  has been widely used to this end despite not being the right criterion for dependent errors. We use the generalized correlated cross-validation (GCCV) as an alternative.  This suggested criterion  is an extension to GCV within the context of correlated errors proposed by \cite{Carmack2012}. It is defined as follows: %by \cite{Carmack2012} is an extension to GCV in the context of correlated errors, which is defined as follows:

%The GLS criterion can be employed to jointly estimate all the parameters associated with the model: $K_x$, $K_\beta$, $\mat{b}$ and $\theta$ although the first two parameters related with the basis for $\mcal{X}$ and $\beta$ are typically chosen \emph{a priori} taking into account, for instance, the quality of the data and its representation on the discretization grid or other considerations related with the data generating process (smoothness, physical restrictions, interpretability,...). Finally, the GLS criterion can be expressed as:\label{metodos}
%$$\mathop{min}_{K_x, K_\beta,\mat{b},\theta} \mathop{GLS}=\mathop{min}_{ \mat{b},\theta}\lrp{y-\mat{Z}\mat{b}}^\prime\Sigma(\theta)^{-1}\lrp{y-\mat{Z}\mat{b}}$$

%The previous equation, mostly depending on the structure of $\Sigma$ could be hard to minimize or computationally expensive.  So, we simplify it assuming $K_x=K_\beta$ and select the number of basis components using an proper validation criteria. Generalized cross-validation (GCV) has been  widely used, and penalizes the fit by the trace of the smoother matrix assuming independent errors. The generalized correlated cross--validation (GCCV) criterion proposed by \cite{Carmack2012} is an extension to GCV in the context of correlated errors, which is defined as follows:
\[
%GCV(K_\beta\gamma,\theta) = \lrp{\frac{y_{i}-\hat{y}_{i,\gamma}}{\sqrt{\varXi_i}}}^\prime\Sigma(\theta)^{-1}\lrp{\frac{y_{i}-\hat{y}_{i,\lrp{ h }}}{\sqrt{\varXi_i}}},
GCCV(K_x,K_\beta,\mat{b},\theta)=\frac{\sum_{i=1}^n \lrp{y_{i}-\hat{y}_{i,\mat{b}}}^2}{\lrp{1-\frac{\mathop{tr}(\mat{C})}{n}}^2} 
\]
where $\mat{C}=2\mat{S\Sigma(\theta)}-\mat{S\Sigma(\theta)S^\prime}$ takes into account the effect of the dependence, the trace of $\mat{C}$ is an estimation of the degrees of freedom consumed by the model and $\mat{S}$ is the smoother matrix. 

The important advantage of this criterion is that it is rather easy to compute. It avoids the need to compute the inverse of the matrix $\Sigma$.  Even so, the GLS criterion mostly depends on the structure of $\Sigma$ and could be hard to minimize or computationally expensive. In these cases, the following iterative procedure (iGLS) is equivalent to the classical GLS.  

\begin{enumerate}[(1)]
	\item Begin with a preliminary estimation of $\hat{\theta}=\theta_0$ (for instance, $\theta_0=0$). Compute $\hat{W}$.
	
	\item Estimate $\mat{b}_\Sigma=\mat{(Z^\prime\hat{W}Z)^{-1}Z^\prime\hat{W}}y$
	
	\item Based on the residuals, $\hat{e}=\lrp{y-\mat{Z}\mat{b}_\Sigma}$, update $\hat{\theta}=\rho\lrp{\hat{e}}$ where $\rho$ depends on the dependence structure chosen.
	
	\item Repeat steps 2 and 3 until convergence (small changes in $\mat{b}_\Sigma$ and/or $\hat{\theta}$) 
\end{enumerate}

%--------------------

\section{Simulation}\label{simulation}
We used two functional linear models (FLM) included in \cite{Cardot2003} to compare the effect of a temporal dependence. We specifically generate samples of size $n=100$ from the FLM model  $y=\inprod{\mcal{X}}{\beta}+\epsilon$, $\mcal{X}$ being a Wiener process defined on $\lrc{0,1}$ and $\epsilon$ an AR(1) process with parameter $\phi$ and variance $\V{\epsilon}=snr$\footnote{Signal to noise ratio}$\V{\inprod{\mcal{X}}{\beta}}$. The two models differ only in the $\beta$ parameter that are respectively: 
 
 \begin{enumerate}
 	\item[(a)] $\beta(t)=2\sin(0.5\pi t)+4\sin(1.5\pi t)+5\sin(2.5\pi t), t \in\left[0,1\right],$ \\
 	\item[(b)] $\beta(t)=\log(15t^2+10)+\cos(4\pi t), t\in\left[0,1\right].$  
 \end{enumerate}
 
 The scenario (a) corresponds to a $\beta$ parameter which has an exact representation respect to the first three theoretical principal components (PC) of the Wiener process. On the contrary, the $\beta$ parameter for scenario (b) cannot be well represented using a small number of PCs of the Wiener process. In both cases, we used principal components and B-splines to estimate the  $\beta$ parameter  and we employed the same basis selected for $\beta$ to represent $\mcal{X}$.    
% let $\beta(t)=\sqrt{2}\nu_1+2\{2}\nu_2+(5/\sqrt{2})\nu_3$

%-------------------------------

%For sake of simplicity, we only show here the results for model (a). The results for model (b) can be revised in the appendix. %The results for  the model (b)  are in Appendix 2.
\setlength{\tabcolsep}{0.9mm}
\begin{scriptsize}
	\begin{table}[htb]
		%$$\phi$$
		\centering
		\begin{tabular}{r|cccc|cccc}
			%      &  & $\phi$, PC & &  &  & $\phi$, BSP & &  \\   
			\multicolumn{1}{c|}{} & \multicolumn{4}{c|}{ $\phi$, PC}& \multicolumn{4}{c}{$\phi$, BSP } \\ \hline
			$snr$& 0.0 & 0.3 & 0.6 & 0.9 & 0.0 & 0.3 & 0.6 & 0.9 \\ 
			\hline
			0.05 & 3.76 & 3.85 & 3.99 & 3.69 & 6.92 & 6.87 & 7.09 & 6.34 \\ 
			0.10 & 3.69 & 3.63 & 3.87 & 3.63 & 6.54 & 6.71 & 6.88 & 6.01 \\ 
			0.20 & 3.51 & 3.52 & 3.66 & 3.58 & 6.29 & 6.18 & 6.53 & 5.79 \\   \hline 
			\hline 
		\end{tabular}
		\caption{Average of number of basis elements selected by GCCV criterion.}
		\label{tab:kmax}
	\end{table}
\end{scriptsize}

Tables~\ref{tab:kmax} to \ref{tab:MSE} summarizes the results for the first model (a)  to show, respectively, the average number of selected components chosen using GCCV criterion, the mean square error (MSE) for estimation of $\beta$, the MSE for estimation of $\phi$ and the mean square prediction errors (MSPE) for horizons $1, 5$ and $10$ for the $B=1,000$ replicas. In these results, LM denotes the estimation though a classical functional linear model whereas GLS and iGLS correspond, respectively, to the functional GLS and functional iGLS methods  (shown in Section~\ref{methods})  for AR(1) dependent errors.
%from a first-order autoregressive process 
%The number of components using B-splines 

\setlength{\tabcolsep}{0.9mm}
\begin{scriptsize}
	\begin{table}[!htb]
		$$\E{\norm{\beta-\hat{\beta}}^2}$$
		\centering
		\begin{tabular}{ll|cccc|cccc}
			\hline
			% &  &  &  &$\phi$, {PC} &  &  & $\phi$, {BSP} &  & \\  \hline
			\multicolumn{2}{c|}{} & \multicolumn{4}{c|}{ $\phi$, PC}& \multicolumn{4}{c}{$\phi$, BSP } \\ \hline
			$snr$ & Model & 0 & 0.3 & 0.6 & 0.9  & 0 & 0.3 & 0.6 & 0.9 \\ 
			\hline
			0.05 & LM & 0.51 & 0.50 & 0.50 & 0.49 & 0.97 & 0.92 & 0.97 & 0.89 \\ 
			0.05 & GLS & 0.51 & 0.48 & 0.44 & 0.46 & 0.97 & 0.89 & 0.67 & 0.50 \\ 
			0.05 & iGLS & 0.51 & 0.48 & 0.45 & 0.47 & 0.97 & 0.86 & 0.66 & 0.49 \\  \hline
			0.10 & LM & 0.60 & 0.61 & 0.59 & 0.57 & 1.38 & 1.35 & 1.35 & 1.21 \\ 
			0.10 & GLS & 0.61 & 0.58 & 0.50 & 0.49 & 1.39 & 1.27 & 1.00 & 0.66 \\ 
			0.10 & iGLS & 0.61 & 0.58 & 0.51 & 0.50 & 1.38 & 1.23 & 0.93 & 0.67 \\  \hline
			0.20 & LM & 0.76 & 0.76 & 0.73 & 0.75 & 1.79 & 1.70 & 1.90 & 1.75 \\ 
			0.20 & GLS & 0.77 & 0.72 & 0.63 & 0.58 & 1.80 & 1.62 & 1.35 & 0.80 \\ 
			0.20 & iGLS & 0.77 & 0.71 & 0.61 & 0.56 & 1.80 & 1.58 & 1.30 & 0.78 \\  \hline
			\hline
		\end{tabular}
		\caption{ Mean square error of $\beta$ parameter. Model (a)}
		\label{tab:beta}
	\end{table}
\end{scriptsize}

We use the GCCV criterion to determine the number of FPC with a search range from 1 to 8. Table~\ref{tab:kmax} shows an average number  of FPC selected components between $3$ and $4$ with a slight tendency to lower values as the $snr$ grows. The range for components using cubic B-splines is from $4$ to $11$, even though we have no theoretical quantity to compare it to, the average number of selected components is between 6 and 7. It seems that there are no trends with respect to the $\phi$ values. Table~\ref{tab:beta} clearly shows the advantage of the PC estimator over the B-splines. The estimation error using B-splines typically doubles the error using PCs. %The advantage that the PC estimator has over the B-splines is clearly shown in Table~\ref{tab:beta} where the estimation error using B-splines typically doubles the error using PCs. 

% latex table generated in R 3.1.2 by xtable 1.7-3 package
% Thu Jan 08 17:26:20 2015
\setlength{\tabcolsep}{0.9mm}
\begin{scriptsize}
	\begin{table}[!htb]
		$$\E{\lrp{\phi-\hat{\phi}}^2}$$
		\centering
		\begin{tabular}{ll|llll|llll}
			\hline
			% &  &  &  & $\phi$, {PC} &  &  & $\phi$, {BSP} &  & \\  \hline
			\multicolumn{2}{c|}{} & \multicolumn{4}{c|}{ $\phi$, PC}& \multicolumn{4}{c}{$\phi$, BSP } \\ \hline
			$snr$ & Model & 0 & 0.3 & 0.6 & 0.9  & 0 & 0.3 & 0.6 & 0.9 \\ 
			\hline
			0.05 & GLS & 0.005 & 0.005 & 0.003 & 0.002 & 0.006 & 0.005 & 0.003 & 0.001 \\ 
			0.05 & iGLS & 0.006 & 0.005 & 0.003 & 0.002 & 0.006 & 0.005 & 0.003 & 0.001 \\ \hline
			0.10 & GLS & 0.005 & 0.004 & 0.004 & 0.002 & 0.005 & 0.004 & 0.004 & 0.001 \\ 
			0.10 & iGLS & 0.005 & 0.004 & 0.004 & 0.002 & 0.006 & 0.004 & 0.004 & 0.002 \\ \hline
			0.20 & GLS & 0.006 & 0.004 & 0.003 & 0.002 & 0.006 & 0.004 & 0.003 & 0.001 \\ 
			0.20 & iGLS & 0.006 & 0.004 & 0.003 & 0.002 & 0.006 & 0.004 & 0.003 & 0.002 \\ 
			\hline \hline
		\end{tabular}
		\caption{Mean square error of $\phi$ parameter. Model (a)}
		\label{tab:phi}
	\end{table}
\end{scriptsize}

\setlength{\tabcolsep}{0.9mm}
\begin{scriptsize}
	\begin{table}[!htb]
		$$MSPE=\sum_{i=1}^{B}\left( y_{n+h}-\hat{y}_{n+h} \right)^2$$
		\centering
		\begin{tabular}{lll|ccc|ccc|ccc|ccc}
			\hline
			% &  $\phi$&&  & 0 &  &  & 0.3 &  &  & 0.6 &  &  & 0.9 & \\  \hline
			&  &$\phi$&  & 0 &  &  & 0.3 &  &  & 0.6 &  &  & 0.9 & \\  \hline
			$snr$ & Model& $h$ & 1 & 5 & 10 & 1 & 5 & 10 & 1 & 5 & 10 & 1 & 5 & 10 \\ 
			\hline  
			0.05 & LM.PC & & 0.07 & 0.07 & 0.07 & 0.07 & 0.08 & 0.07 & 0.09 & 0.08 & 0.08 & 0.07 & 0.07 & 0.08 \\ 
			0.05 & GLS.PC && 0.07 & 0.07 & 0.07 & 0.07 & 0.08 & 0.07 & 0.05 & 0.07 & 0.07 & 0.02 & 0.05 & 0.07 \\ 
			0.05 & iGLS.PC && 0.07 & 0.07 & 0.07 & 0.07 & 0.08 & 0.07 & 0.05 & 0.07 & 0.07 & 0.02 & 0.05 & 0.07 \\  \hline
			0.05 & LM.BSP &&0.07 & 0.08 & 0.07 & 0.07 & 0.08 & 0.07 & 0.09 & 0.08 & 0.08 & 0.07 & 0.07 & 0.08 \\ 
			0.05 & GLS.BSP && 0.07 & 0.08 & 0.08 & 0.07 & 0.08 & 0.07 & 0.05 & 0.07 & 0.08 & 0.01 & 0.05 & 0.07 \\ 
			0.05 & iGLS.BSP &&0.07 & 0.08 & 0.08 & 0.07 & 0.08 & 0.07 & 0.05 & 0.07 & 0.08 & 0.01 & 0.05 & 0.07 \\  \hline \hline
			0.10 & LM.PC &&0.18 & 0.18 & 0.15 & 0.16 & 0.16 & 0.16 & 0.17 & 0.17 & 0.18 & 0.15 & 0.16 & 0.18 \\ 
			0.10 & GLS.PC && 0.18 & 0.18 & 0.15 & 0.14 & 0.16 & 0.16 & 0.11 & 0.16 & 0.17 & 0.04 & 0.12 & 0.17 \\ 
			0.10 & iGLS.PC && 0.18 & 0.18 & 0.15 & 0.14 & 0.16 & 0.16 & 0.11 & 0.16 & 0.17 & 0.04 & 0.11 & 0.16 \\  \hline
			0.10 & LM.BSP && 0.18 & 0.18 & 0.15 & 0.16 & 0.16 & 0.16 & 0.17 & 0.17 & 0.18 & 0.16 & 0.17 & 0.18 \\ 
			0.10 & GLS.BSP && 0.18 & 0.18 & 0.15 & 0.15 & 0.16 & 0.15 & 0.12 & 0.16 & 0.18 & 0.04 & 0.12 & 0.16 \\ 
			0.10 & iGLS.BSP &&0.18 & 0.18 & 0.15 & 0.15 & 0.16 & 0.15 & 0.12 & 0.16 & 0.18 & 0.04 & 0.11 & 0.16 \\  \hline \hline
			0.20 & LM.PC && 0.33 & 0.40 & 0.38 & 0.37 & 0.35 & 0.34 & 0.36 & 0.35 & 0.39 & 0.37 & 0.38 & 0.39 \\ 
			0.20 & GLS.PC &&0.32 & 0.40 & 0.38 & 0.34 & 0.35 & 0.34 & 0.24 & 0.34 & 0.38 & 0.07 & 0.24 & 0.32 \\ 
			0.20 & iGLS.PC& & 0.32 & 0.40 & 0.38 & 0.34 & 0.35 & 0.34 & 0.24 & 0.34 & 0.38 & 0.07 & 0.24 & 0.32 \\  \hline
			0.20 & LM.BSP && 0.34 & 0.41 & 0.39 & 0.38 & 0.38 & 0.34 & 0.38 & 0.37 & 0.40 & 0.38 & 0.39 & 0.40 \\ 
			0.20 & GLS.BSP &&  0.34 & 0.42 & 0.39 & 0.35 & 0.38 & 0.34 & 0.25 & 0.34 & 0.39 & 0.08 & 0.25 & 0.33 \\
			0.20 & iGLS.BSP && 0.34 & 0.42 & 0.39 & 0.35 & 0.38 & 0.34 & 0.25 & 0.34 & 0.39 & 0.08 & 0.25 & 0.32 \\   
			\hline   \hline
		\end{tabular}
		\caption{ Mean square prediction errors for different lags.}
		\label{tab:MSE}
	\end{table}
\end{scriptsize}

In this table, we can also see the improved estimates of the GLS and IGLS method over the LM, especially when $\phi$ grows. The same equivalence is shown in Table~\ref{tab:phi} for the mean square error (MSE) of the $\phi$ parameter, which shows better results as the dependence grows. Finally, Table~\ref{tab:MSE} shows the mean square prediction errors (MSPE) for different lags showing a clear improvement of GLS procedures, specially for large $\phi$ and shorter lags. With respect to the prediction ability between PC or B--splines, the results show that both methods are almost equivalent with minor differences along the table.

%---------------------------------------
\section{Real example: flu prediction}\label{example}
%Influenza is an infectious disease with person--to--person transmission that characteristically occurs as epidemics affecting the whole population (\cite{Watson2012}). The epidemiology of inter-pandemic influenza (also named seasonal influenza) is characterised in temperate zones by epidemics of variable size which occur during the colder winter months (November to April in the Northern Hemisphere and May to September in the Southern hemisphere), each typically lasting $8-10$ weeks (\cite{VanTam2013}). 

%The reasons for the seasonal presentation of epidemics are not entirely clear but might be the result of more favourable environmental conditions for virus survival (\cite{Schaffer1976}). In temperate zones, various theories have been advanced to explain this pattern, including improved virus survival in low temperatures, low humidity and low levels of ultraviolet radiation (\cite{VanTam2013}). %Surveillance systems require accurate indicators that detect possible epidemics in advance. Seasonal epidemics of incfluenza are a problem that concerns public health professionals worldwide, due to the high levels of mortality and morbidity and lost work days.

%\subsection{Data description and transformation}

Galicia is a region of $29,574$ km$^2$ located in Northwest Spain with a population of $2.8$ million people. We analyzed the weekly incidence of reported cases of influenza in Galicia between $2001$ and $2011$ for each of $53$ Galician counties.
$$\mbox{Rate}_{n,s}=\log(cases_{n,s}\times100000/pop_{n,s})$$ for county  $s$ and  week $n$.  The Statistical Institute of Galicia (IGE, {\tt{http://www.ige.eu}}) provided population ($pop$) and the Health Department of Galicia ({\tt{www.sergas.es}}) provided the number of cases of influenza ($cases$).

\begin{figure}[htb]
	\begin{center}
		\includegraphics[width=15cm]{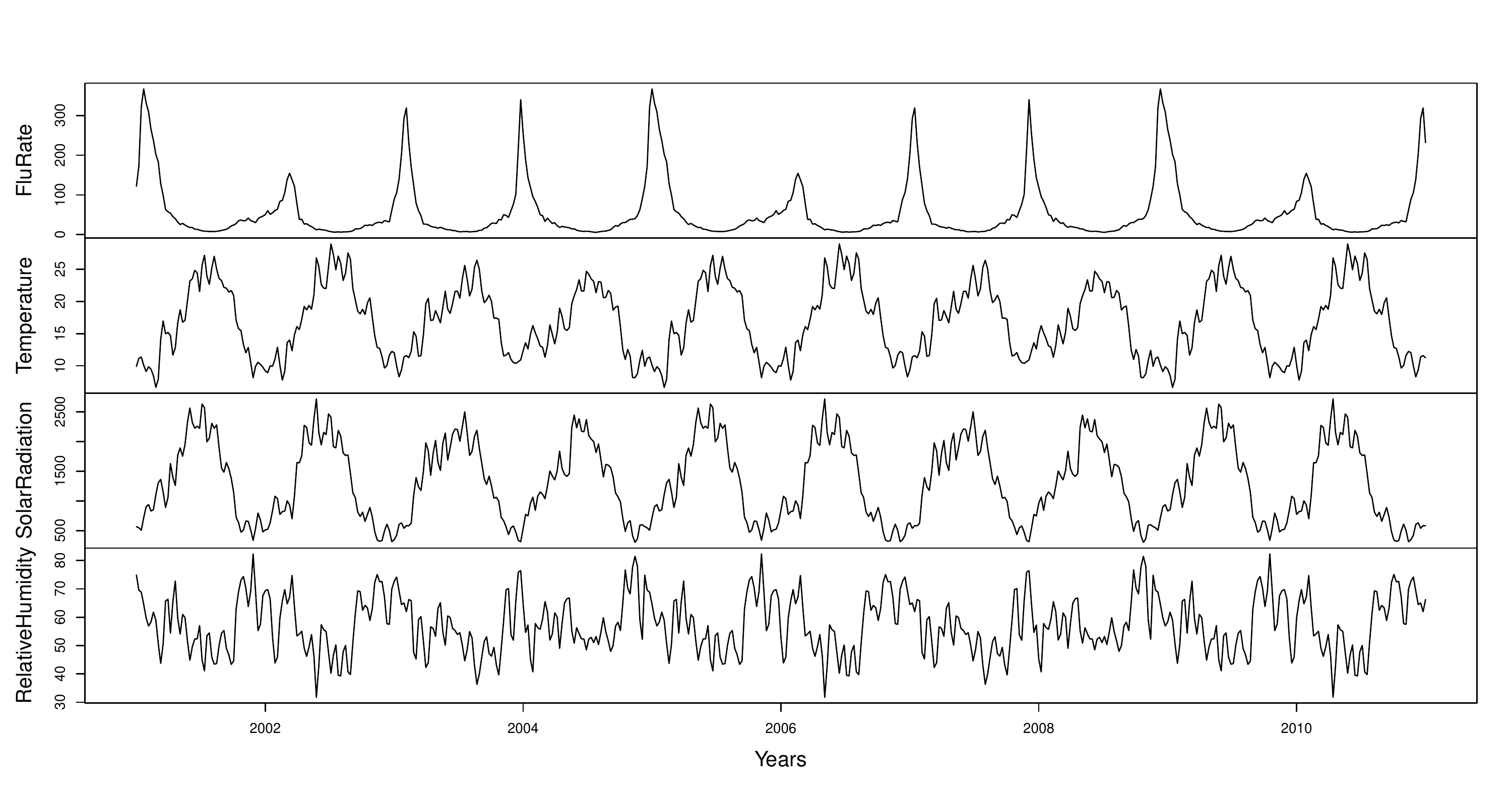}
		\label{figTS}
		\caption{From top to bottom: Weekly influenza rate, and daily average temperature, solar radiation and relative humidity in the Galician region during the period.}
	\end{center}
\end{figure}

The influenza season in Galicia begins in week $40$ and ends in week $20$ of the following year. The goal is to predict the incidence of influenza for the following two weeks ($n+1$ and $n+2$) in the $s$ regions with the available information:

\begin{itemize}
	\item $\mathop{\mbox{Rate}}_{n,s}(w)$: Weekly influenza rate for last 13 weeks, $w\in \lrc{n-12,n}$.	
	%  	=\lrc{\mathop{\mbox{Rate}}_s(t+h)}, h\in \lrc{-12,0}$.
	\item  $\mathop{\mbox{Temp}}_{n,s}(t)$: Daily temperature in Celsius degrees ($^{\circ}$C) for last 14 days, $t\in \lrc{n-i/7,n}$, for  $i=14,\ldots,1$.	
	
	%	\item $\mathop{\mbox{Temp.thres}}_s(t)$: Daily threshold temperature below 10 C during last 14 days as proposed in \cite{Dushoff2006}, $t\in \lrc{n-i/7,n}, for\  i=14,\ldots,0 $.
	
	\item  \cite{Dushoff2006} defined cold as the number of degrees below a threshold temperature:  
	$  \mathop{\mbox{Temp.thres}}_{n,s} = min(\mathop{\mbox{Temp}}_{n,s} - \mathop{\mbox{thres}}  , 0)$ with $\mathop{\mbox{thres}}=10^{\circ} C$. The functional variable is defined as: $\mathop{\mbox{Temp.thres}}_{n,s}(t)$ with $t\in \lrc{n-i/7,n}$, for  $i=14,\ldots,1$.
	%Cold was measured on a monthly basis as the average number of degrees below a threshold temperature (Tthresh) or cold (C) = min(Tthresh ? T, 0), where T is the mean monthly temperature. Cold was measured for each flu year as the average of monthly cold values from November through April.

	\item $\mathop{\mbox{SR}}_{n,s}(t)$: Daily solar radiation ($W/m^2$) for the last 14 days, $t\in \lrc{n-i/7,n}$, for  $i=14,\ldots,1$.
	
	\item $\mathop{\mbox{Hum}}_{n,s}(t)$: Relative humidity for the last 14 days: , $t\in \lrc{n-i/7,n}$, for  $i=14,\ldots,1$.
\end{itemize}

Figure 1 shows that  influenza normally starts in the late autumn and reaches a peak at the beginning of the calendar year. These plots clearly show the large difference between reported influenza cases in winter and summer. We downloaded meteorological data from the regional Weather Service of Galicia ({\tt{http://www.meteogalicia.es/}}).

%---------------------
\subsection{Variable Selection  using Distance Correlation Measure}

Distance correlation $\mathcal{R}$ is a measure of dependence between random vectors introduced by \cite{Szekely2007}. The distance correlation satisfies $0<\mathcal{R}(X,Y)<1$ and its interpretation is similar to the squared Pearson's correlation. However, the advantages of distance correlation over the Pearson correlation is that it defines $\mathcal{R}(X,Y)$ in arbitrary finite dimensions of $X$ and $Y$ and $\mathcal{R}$ characterises independence, i.e. $\mathcal{R}(X,Y)=0\Leftrightarrow X,Y$ are independent. Recently, \cite{Lyons2013} provided conditions for the application of the distance correlation to functional spaces.  So, this measure seems to be a good indicator of the correlations between functional and multivariate variables that may be useful for designing a functional linear model (for instance, avoiding  variates with high collinearity). We can easily calculate the empirical distance correlation ${\mathcal{R}_{n,s}(\mathbf{X,Y})}$ as
$$
\mathcal{R}_{n,s}(\mathbf{X,Y})=\sqrt{ \frac {\mathcal{V}^2_{n,s}(\mathbf{X,Y})}
	{\sqrt{ \mathcal{V}^2_{n,s} (\mathbf{X}) \mathcal{V}^2_{n,s}(\mathbf{Y})}}}.
$$		
where ${\mathcal{V}_{n,s}(\mathbf{X,Y})}$ is the empirical distance covariance defined by
$  \mathcal{V}^2_{n,s} (\mathbf{X,Y}) = \frac{1}{n^2} \sum_{k,\,l=1}^n A_{kl}B_{kl} $
where $A_{kl} = a_{kl}-\bar a_{k.}- \bar a_{.l} + \bar a_{..}$  and $ B_{kl} = b_{kl}-\bar b_{k.}- \bar b_{.l} + \bar b_{..}.$ with
$ a_{kl} = \|X_k - X_l\|, \quad b_{kl} = \|Y_k - Y_l\|, \quad k,l=1,\dots,n,$
and the subscript ${.}$ denotes that the mean is computed for the index that it replaces.
% and $\| \  \|_p,$ denotes the p--norm. 
Similarly,  ${\mathcal{V}_{n,s}(\mathbf{X})}$  is the non-negative number defined by $ {\mathcal{V}^2_{n,s}(\mathbf{X})}= \mathcal{V}^2_{n,s} (\mathbf{X,X}) = \frac{1}{n^2} \sum_{k,\,l=1}^n A_{kl}^2 $.

We used the distance correlation measure  $\mathcal{R}$ to select the information relevant to the prediction of influenza rate. The results are shown in Table~\ref{tab:dcc}.  Relative humidity $\mathop{\mbox{Hum}}_{n,s}(t)$ is the variable that has the lowest correlation with the influenza rate  $\left\{\mathop{\mbox{Rate}}_{n+1,s},\mathop{\mbox{Rate}}_{n+2,s}\right\}$  and therefore the set of possible predictor variables is reduced to: % the next subset:
 $\left\{
\mathop{\mbox{Rate}}_{n,s}(w), \mathop{\mbox{Temp}}_{n,s}(t), \mathop{\mbox{Temp.thres}}_{n,s}(t),\ \mathop{\mbox{SR}}_{n,s}(t)
\right\}$. Besides, the distance correlation values lead to designing the models avoiding closely related covariates closely related (for instance, $\mathop{\mbox{Temp}}_{n,s}(t)$ and $\mathop{\mbox{Temp.thres}}_{n,s}(t)$)  and thereby reduces the number of possible models tested.
%orden previo (ranking incial de las variables predictoras)
\begin{table}[!htb]
	\centering
	\footnotesize{
		\begin{tabular}{l|ccccc|cc}
			\hline
			$\mcal{R}$ &  $\mathop{\mbox{Rate}}_{n,s}(w)$ & $\mathop{\mbox{Temp}}_{n,s}(t)$ & $\mathop{\mbox{Temp.thres}}_{n,s}(t)$& $\mathop{\mbox{SR}}_{n,s}(t)$ & $\mathop{\mbox{Hum}}_{n,s}(t)$  & $\mathop{\mbox{Rate}}_{n+1,s}$ & $\mathop{\mbox{Rate}}_{n+2,s}$   \\
			\hline
			$\mathop{\mbox{Rate}}_{n,s}(w)$ & 1.00  &  0.49 &    0.42 &   0.40 & 0.24&  0.67 &   0.61\\
			$\mathop{\mbox{Temp}}_{n,s}(t)$ & 0.49  &  1.00 &    0.89 &   0.82 &   0.53& 0.49&    0.46\\
			$\mathop{\mbox{Temp.thres}}_{n,s}(t)$& 0.42 &   0.89  &   1.00 &   0.72 & 0.43&  0.41&    0.41\\
			$\mathop{\mbox{SR}}_{n,s}(t)$ &  0.40 &   0.82&     0.72&    1.00&    0.70&0.50&    0.50\\
			$\mathop{\mbox{Hum}}_{n,s}(t)$  & 0.24 &   0.53&     0.43&    0.70&   1.00& 0.29&    0.27\\
			\hline
		\end{tabular}
		\caption{Distance correlation $\mathcal R$ between the response at week ${n+1}$ and ${n+2}$ and functional covariates at week ${n}$.}
		\label{tab:dcc}
		}

\end{table}

%---------------------
\subsection{Prediction using temporal dependence structure}

%For ease of simplicity, the GLS procedure was employed to model the last two years (107 weeks) in the 53 counties using a simple autoregressive structure of order one AR(1) and therefore the elements $i,j$ of the correlation matrix will be composed by  $\Sigma_{i,j}=\sigma\phi^{\left|i-j\right|}$.  

%The results between the three proposed models were similar with a slightly advantage in model (c). However, the functional variate solar radiation $\mbox{SR}$ is often hard to obtain and depends on specialised devices whereas the covariates related with temperature are easily available with quite standard equipment. Indeed, between the two temperature related variables, the $\beta$ parameter associated with the $\mbox{Temp.thres}$ has a nicer interpretation than the $\mbox{Temp}$ itself. 

We employed a rolling analysis to compare the models in a predictive scenario.
%To compare the models in a predictive scenario, we have employed a rolling analysis. 
In our case, we initially split the data into an estimate sample of length $j=1,\ldots,n=104$ weeks  (2 years) in $s=53$ counties  and evaluated a predictive sample size in next two weeks $n+1$ and $n+2$ in the $s=53$ counties. The rolling was performed along the epidemic periods for 2 years ($J=40$ weeks) by computing the mean square prediction error: $MSPE=\frac{1}{n+J}\sum_{j=n+1}^{n+J}\frac{1}{s}\sum_{r=1}^{s} {\left({Rate}_{j,r}-\widehat{Rate}_{j,r}\right)^2}$. For ease of simplicity,  we employed the functional GLS procedure was employed simple autoregressive structure of order one AR(1). Therefore the elements $k,l$ of the correlation matrix consists in $\Sigma_{k,l}=\sigma\phi^{\left|k-l\right|}$.

\begin{table}
	\centering
	%	\footnotesize{
	\small{
		\begin{tabular}{l|l||rr||rr}
			% \hline
			% & &     $\hat{\phi}$&& df &(residual) & R^2 (adj)&& $GCCV(\phi)$& \\ 
			\multicolumn{2}{c||}{} & \multicolumn{2}{c||}{$n+1$}& \multicolumn{2}{c}{$n+2$}\\ \hline
			%			\multicolumn{2}{l||}{} &  \multicolumn{2}{c|}{$Var(\hat{\varepsilon})$} &  \multicolumn{2}{c||}{MSPE} & \multicolumn{2}{c|}{$\hat{\phi}$}  &  \multicolumn{2}{c|}{$Var(\hat{\varepsilon})$} &  \multicolumn{2}{c}{MSPE}   \\
			Model& Covariates & FLM & FGLS & FLM & FGLS \\  
			\hline
			(a)&  $\mbox{Rate}_{n,s}(w)$&  0.56 & 0.49 &  0.86 & 0.83     \\  
			(b)&  $\mbox{Temp}_{n,s}(t)$& 1.96 & \textbf{0.46} &  2.02 & 0.79\\ 
			(c)&  $\mbox{Temp.thres}_{n,s}(t)$& 2.57 &  0.50 & 2.53 & 0.80  \\ 
			(d)&  $\mbox{SR}_{n,s}(t)$& 1.48 &  \textbf{0.46} &  1.58 &  \textbf{0.77}\\ 
			(e)&  $\mbox{Rate}_{n,s}(w),\mbox{Temp}_{n,s}(t)$& 0.64 &  0.53 &   0.91 & 0.86\\  
			(f)&  $\mbox{Rate}_{n,s}(w),\mbox{Temp.thres}_{n,s}(t)$& 0.67 &  0.57&  0.92&  0.87\\ 
			(g)&  $\mbox{Rate}_{n,s}(w),\mbox{SR}_{n,s}(t)$& 0.57&  0.50& 0.85& 0.84 \\ 
			(h)&  $\mbox{Temp}_{n,s}(t),\mbox{SR}_{n,s}(t)$& 1.85&  0.48& 1.93& 0.83 \\ \hline			
			\hline 
		\end{tabular}
		\caption{Serial prediction errors (MSPE) for influenza seasonal period using the rolling procedure.}
		\label{tab:mse3}
	}
\end{table}%all: all weeks, Epidemic: week 40 to 20, Flu season: weeks with rate>100. 

% % % % % % % % % % % % % % % % redactar
Table~\ref{tab:mse3} summarises the predictive errors (MSPE) for the influenza season. 
%The estimate of the predictor  $\mathop{\mbox{Rate}}_{n,s}(w)$, in a way, is  similar to the estimation of the  dependence structure of the residuals using an autocorrelation function. That is why the models with the predictor $\mathop{\mbox{Rate}}_{n,s}(w)$, models (a), (e), (f) and (g), the gain with respect to the MSPE in the dependent errors estimation (FGLS model) is relatively small with respect to considering independent errors (FLM). 
The gain of the FGLS, in terms of  MSPE, with the predictor $\mathop{\mbox{Rate}}_{n,s}(w)$  (models (a), (e), (f) and (g)) is relatively small with respect to considering FLM models. However, meteorological models such as models (b), (c), (d) and (h), the GLS setting reduces prediction error, at least, to one half. In the FGLS estimation the serial dependence considered  is a quite simple AR(1). The $\hat{\phi}$ parameter ranges  for $0.85$ to $0.9$ for meteorological models against   $0.5$ to $0.55$ for models including the Rate as a covariate. The latter case is  expected because the $\mathop{\mbox{Rate}}_{n,s}(w)$ accounts for part of the temporal dependence. The best models (b) and (d) with FGLS show small differences between them. The model (h) does not improve the results of models (b) and (d) in terms of MSPE. In fact it worsens them and this is probably due to collinearity. % between these variables thus corroborating the interpretation  based  on the correlation distances measure, see table~\ref{tab:dcc}.  
Among the models (b) and (d), the first is preferable because it is easier to interpret.  Besides, solar radiation is often hard to obtain and depends on specialised devices whereas the covariates related to temperature are readily available with quite standard equipment.

\begin{figure}[!h]
	\begin{center}
		\includegraphics[width=11cm]{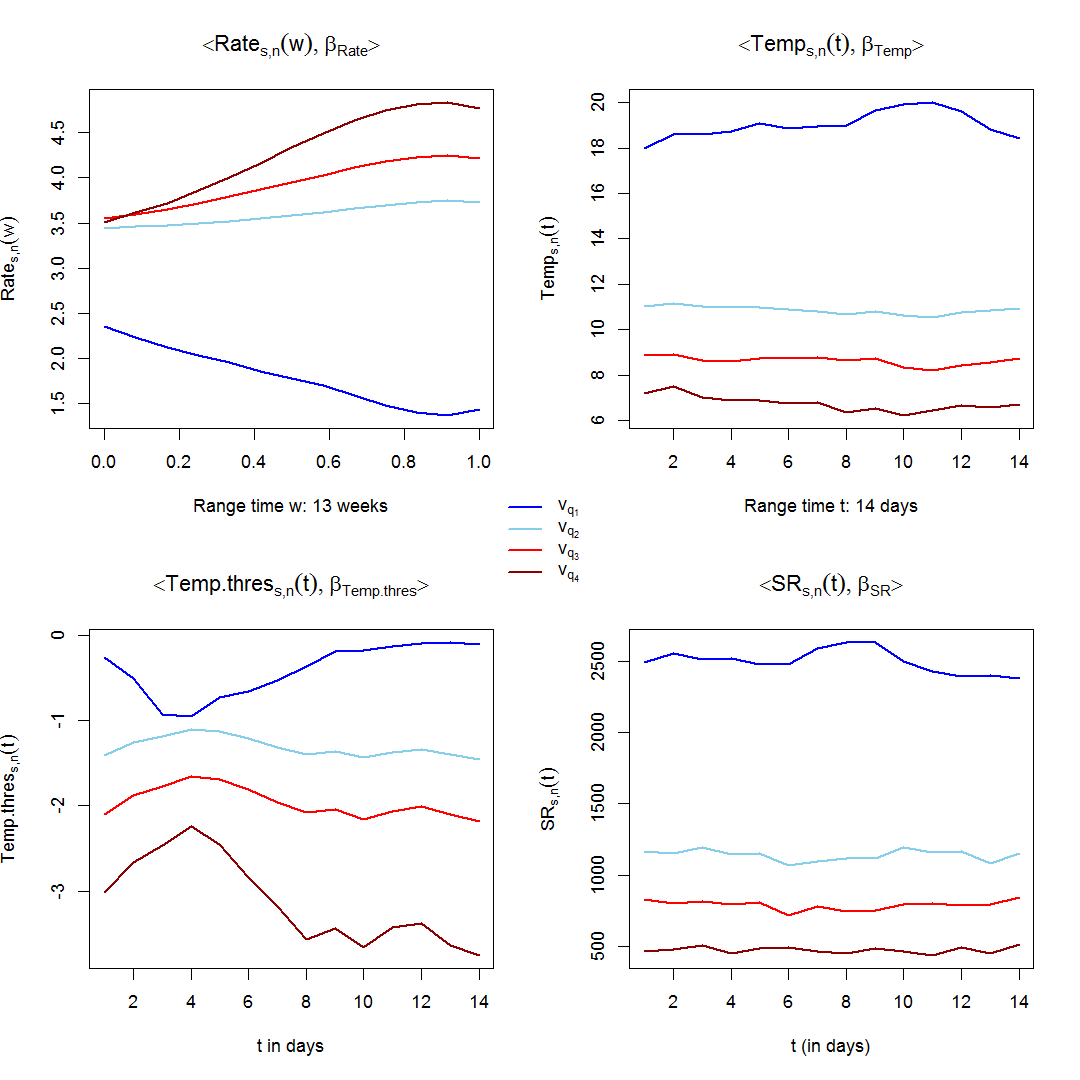}
		\caption{
			Shape of rate curves (on left) and temperature threshold curves (on right) in function  of their projection value $v_X=\left\langle X,\hat{\beta}\right\rangle$.  Average of curves: in $1$st quartile of $v_x$ (dark blue line), in $2$nd quartile of $v_x$ (blue line), in $3$rd quartile of $v_x$ (red line) and in $4$th quartile of $v_x$ (dark red line).		}
		\label{fig6}
	\end{center}
\end{figure}

Indeed, between the two temperature-related variables, the $\hat{\beta}$ parameter associated with the temperature  $\mathop{\mbox{Temp}}_{n,s}(t)$ has a nicer interpretation. % than the $\mbox{Temp}$ itself. 
To this end, we have computed for models (a), (b) (c) and (d), the quantities  $v_i=\left\langle \mcal{X}_i,\hat{\beta}\right\rangle$ that include the contribution of each variable to the prediction of the influenza rate. We used these contributions to determine the pattern of curves that most influenced the increase and decrease in the incidence rate. Figure~\ref{fig6} shows the pattern of curves that most contributed to increasing (in red scale) and decreasing (in blue scale) the influenza rate. In particular, we split the data with respect to the quartiles of $v_i$ and computed the average functional variate for each group (blue, sky blue, red and dark red, respectively correspond to quantile groups $[0,0.25]$, $[0.25,0.5]$, $[0.5,0.75]$ and $[0.75,1]$). This assesses the evaluation of the contribution of these curves in the response.

As the temperature increases (top left of Figure~\ref{fig6}),  the curves around  $7^{\circ}$C (dark red line correspond  to $v_{q_4}$ which is the highest contribution) provide an increase of the estimated influenza rate. In contrast, the contribution of the curves around $19^{\circ}$C have no impact on the influenza rate (blue line correspond  $v_{q_1}$, low or null contribution). The interpretation of the other models is straightforward.%Incluir descripción

% comentar algo respecto si puede asumirse estructruas similares en las diferentes regiones o no...como es que no tb decir que puede incluirse esa estructura espacial 
%Observations in the 53 counties (groups) were assumed to be independent, with $\Sigma$ being a block diagonal matrix of the correlation matrices per county,  $\Sigma=\sigma^2 diag\left( \Sigma_1(\phi_1),\ldots,\Sigma_s(\phi_s)\right)$, where  $\sigma^2=\sigma^2_1,\ldots,\sigma^2_s$ with $s=1,\ldots,53$. 

%El 	$\sigma_s$ para las  $s=1,\ldots,53$ puede ser hasta más de tres veces el valor del una de las comarcas.  Tanto para el modelo 1 como para el modelo 2.  Esto no hace más que confirma la hipótesis que no puede asumirse la misma $\sigma$.

%En este punto la clave es,  if Observations in the 53 counties (groups) pueden asumir que son independientes, en ese caso $\Sigma$ sería  a block diagonal matrix of the correlation matrices per county,  $\Sigma=\sigma^2 diag\left( \Sigma_1(\phi),\ldots,\Sigma_s(\phi)\right)$, where  $\sigma^2=\sigma^2_1,\ldots,\sigma^2_s$ with $s=1,\ldots,53$. 	

%%%%%%%%%%%%%%%%%%%%%%%%%%%%%%%%%%%%%%%%%%%%%%%%%%%%%%%%%%%%%%%%%%%%%%%%%%

%-----------------
\section{Conclusion}\label{conclusion}
This paper extends the GLS model from a multivariate to a functional framework: it thereby allows us to estimate functional regression models with temporal or spatial covariance errors structure in a simple way. It proposes an iterative version of the GLS estimator, that can help to model very complicated dependence structures. This procedure (called iGLS) is much simpler than GLS in terms of the optimization function to be accomplished but, of course, it may take longer due to the iterations. 
%Even though this procedure (called iGLS) is much simpler than GLS in terms of the optimization function, it may take longer due to the iterations.
However, iGLS may be the only option when the sample size or the dimension of the parameter increases and the joint optimization performed by GLS is not affordable (in terms of complexity or memory consumption).

A simulation study shows that the GLS estimators improve the classical approach because they provide better estimations of the parameters associated with the regression model and extremely good results from the predictive point of view, specially for short lags.

We applied the GLS procedure to the prediction of the influenza rate using readily available functional variables. These kinds of models are extremely useful to health managers in allocating resources in advance for an epidemic outbreak. In particular, the nice interpretation of the model shows that influenza may increase due to a cold wave with daily temperatures of around $7^{\circ}$C for two weeks.

In our example,  the error structure is estimated by a simple AR(1) and it obtains a good fit for time dependence. We have also tried other ARMA models that have rendered similar results. Our method can additionally be used to explore more complex dependence structures like heterogeneous covariances by counties or even spatio--temporal modelling.  %In our case, this fit would entail adding the estimate of $2 \times 53 = 106$ parameters to our final model. %decir q son 53 es el numero de regiones

Our results are consistent with much of the literature on influenza. By simply using the temperature variables that are easy to obtain, the model achieves results akin to those of models with variables that are more difficult to measure.
%The modeling agency gets less informative variables models have the same result as models more complex and more difficult to obtain variables. The model can, for example, using temperature (readily available) instead of include for example the rate (more difficult to achieve).
%The incidence of influenza is not typically measured directly, but rather derived from the incidence of influenza-like illness (ILI) (\cite{vanNort2012}). Weather factors have been shown to influence the manifestation of influenza-like symptoms. \cite{Shoji2011} found that  influenza epidemics always occur in winter in Japan and are associated with  low temperature and low humidity. The relationship between humidity, temperature and influenza mortality occurs not only in the Southeast Asia (\cite{Loh2011}) but also in countries like the United States (\cite{Barreca2012}). \cite{Tamerius2013} found that the relationship between influenza and weather variables occurs in the vast majority of countries around the world. Our model could include more information if this was easily available, such as the age of those infected, vaccination coverage, influenza virus type, absolute humidity or daily on-line influenza data, the consumption of anti-influenza drugs, social networks data and other factors, but this would imply a deeper descriptive analysis of the data and the incorporation of greater complexity in the regression model. %Google flu trends data 
%In this study we used a model of temporal dependence on errors but also developed code also allows estimating models that include spatial dependence.
Our results also show that the temporal dependence of the influenza virus is strong and stable over time. Future studies could extend  generalized mixed models to the functional framework.

\section*{Acknowledgments}\label{Acknowledgments}

This research was partially supported by the Spanish Ministerio de Ciencia y Tecnolog\'\i a, grant  MTM2013-41383-P. The authors thanks the healthcare provider:  the Service  Epidemiology of the Dirección Xeral de Saúde Pública  (SERGAS) from the Consellería de Sanidade (Xunta de Galicia). %the Health Department of Galicia  (SERGAS).

%healthcare provider (Servizio Galego de Saúde, SERGAS). 
% from the Health Department of Galicia ({\tt{www.sergas.es}}).
%health service of Galicia (SERGAS) offering a

\bibliographystyle{spbasic}
\bibliographystyle{spbasic}      % basic style, author-year citations
\bibliographystyle{spmpsci}      % mathematics and physical sciences
\bibliographystyle{spphys}       % APS-like style for physics
\bibliography{ref} 

\newpage
\section*{Appendix A. Source code}\label{source} 

% This section provides information on the procedures used in the study. The supplementary material consists of three R-script files containing source code for computing the functional version of  distance correlation proposed by \cite{Szekely2007} and estimating (and predicting) the functional regression model with correlated errors. \\
% (i) {\tt{dcor.fdist -- R}} code for computing the distance correlation between multivariate and functional objects.\\
% (ii) {\tt{fregre.gls -- R}}, {\tt{fregre.igls -- R}} code for estimating the functional regression model with correlated errors.\\
% (iii) 	{\tt{predict.fregre.gls -- R}}, {\tt{predict.fregre.gls -- R}} code for predicting the functional regression model with correlated errors.

This section provides information on the procedures used in the study.  All functions are included  in version 1.4 of the {fda.usc} package (\cite{Febrero2012}) of software {R}. The function  {\tt{fregre.gls}} (and   {\tt{predict.fregre.gls}}) estimates (and predicts) the functional regression model with correlated errors, function {\tt{fregre.igls}} is an iterative version of the previous one, function  {\tt{dcor.fdist}} computes the distance correlation between multivariate and functional objects and function  {\tt{GCCV.S}}  computes the GCCV criterion.

\section*{Appendix B: Simulation (b)}\label{ap2}
The results for  Simulation (b)  are summarised in Tables~\ref{tab:kmax2}, \ref{tab:beta2},  \ref{tab:phi2}  and \ref{tab:MSE2}. For the sake of simplicity, the results from the iGLS method are not shown because, as in the previous case, the numbers are almost identical with GLS. In this second model, the estimation of $\beta$ cannot be done efficiently with the eigenfunctions of the Wiener process and so, the PC method has no a clear advantage over the B-splines. The number of selected components in Table~\ref{tab:kmax2} is clearly low for PC and quite unstable for higher values of $\phi$. This is also reflected in Table~\ref{tab:beta2} where, especially for $\phi=0.9$, the estimation error is lower for the B-spline procedure. Note that, the estimation error can be split in two parts: a systematic one due to the lack of representation of $\beta$ using a particular basis and the approximation one due to the particular estimation of that basis representation with the data at hand. This also affects the estimation of the dependence parameter as it is shown in Table~\ref{tab:phi2} where the mean square errors provided are larger than in the previous model. In any case, the mean square prediction errors in Table~\ref{tab:MSE2} are better for the GLS procedure than for the classical one.

\setlength{\tabcolsep}{0.9mm}
\begin{scriptsize}
	\begin{table}[!htb]
		%$$\phi$$
		\centering
		\begin{tabular}{r|cccc|cccc}
			%     &  & $\phi$, {PC} & &  &  & $\phi$, {BSP} & &  \\ 
			\multicolumn{1}{c|}{} & \multicolumn{4}{c|}{ $\phi$, PC}& \multicolumn{4}{c}{$\phi$, BSP } \\ \hline
			
			$snr$& 0 & 0.3 & 0.6 & 0.9 & 0 & 0.3 & 0.6 & 0.9 \\ 
			\hline
			0.05 & 2.02 & 2.13 & 2.37 & 4.00 & 5.90 & 5.70& 5.86 & 6.00 \\ 
			0.10  & 1.59 & 1.63 & 1.94 & 3.10 & 5.82 & 5.41& 5.77 & 5.85\\ 
			0.20  & 1.29 & 1.27 & 1.49 & 2.29 & 5.72 & 5.85& 5.80 & 5.79 \\   \hline 
			\hline 
		\end{tabular}
		\caption{Average of number of basis elements selected by GCCV criterion. Model (b).}
		\label{tab:kmax2}
	\end{table}
\end{scriptsize}

\setlength{\tabcolsep}{0.9mm}
\begin{scriptsize}
	\begin{table}[!htb]
		$$\E{\norm{\beta-\hat{\beta}}^2}$$
		\centering
		\begin{tabular}{ll|cccc|cccc}
			\hline
			% &  &  &  &$\phi$, {PC} &  &  & $\phi$, {BSP} &  & \\  \hline
			\multicolumn{2}{c|}{} & \multicolumn{4}{c|}{ $\phi$, PC}& \multicolumn{4}{c}{$\phi$, BSP } \\ \hline
			$snr$ & Model & 0 & 0.3 & 0.6 & 0.9  & 0 & 0.3 & 0.6 & 0.9 \\ 
			\hline
			0.05 & LM & 1.14 & 1.12 & 1.10 & 0.90 & 1.70 & 1.55 & 1.35 & 0.75 \\ 
			0.05 & GLS & 1.14 & 1.11 & 1.07 & 0.88 & 1.71 & 1.42 & 0.94 & 0.40 \\    \hline
			0.10 & LM & 1.19 & 1.18 & 1.16 & 1.02 & 2.50 & 2.26 & 1.93 & 1.02 \\ 
			0.10 & GLS & 1.19 & 1.18 & 1.12 & 0.98 & 2.50 & 2.02 & 1.26 & 0.47 \\    \hline
			0.20 & LM & 1.24 & 1.23 & 1.21 & 1.11 & 3.52 & 3.52 & 2.84 & 1.43 \\ 
			0.20 & GLS & 1.24 & 1.22 & 1.18 & 1.07 & 3.55 & 3.15 & 1.91 & 0.58 \\    \hline
			\hline
		\end{tabular}
		\caption{Mean square error of $\beta$ parameter. Model (b)}
		\label{tab:beta2}
	\end{table}
\end{scriptsize}

\setlength{\tabcolsep}{0.9mm}
\begin{scriptsize}
	\begin{table}[!htb]
		$$\E{\lrp{\phi-\hat{\phi}}^2}$$
		\centering
		\begin{tabular}{ll|cccc|cccc}
			\hline
			% &  &  &  & $\phi$, {PC} &  &  & $\phi$, {BSP} &  & \\  \hline
			\multicolumn{2}{c|}{} & \multicolumn{4}{c|}{ $\phi$, PC}& \multicolumn{4}{c}{$\phi$, BSP } \\ \hline
			$snr$ & Model & 0 & 0.3 & 0.6 & 0.9  & 0 & 0.3 & 0.6 & 0.9 \\ 
			\hline
			0.05 & GLS &  0.011 & 0.011 & 0.011 & 0.019 & 0.012 & 0.011 & 0.007 & 0.005 \\ 
			0.10  & GLS &  0.009 & 0.010 & 0.009 & 0.011 & 0.011 & 0.010 & 0.007 & 0.004  \\ 
			0.20  & GLS & 0.011 & 0.010 & 0.010 & 0.014 & 0.012 & 0.011 & 0.008 & 0.004 \\ 
			\hline \hline
		\end{tabular}
		\caption{Mean square error of $\phi$ parameter. Model (b)}
		\label{tab:phi2}
	\end{table}
\end{scriptsize}

\setlength{\tabcolsep}{0.9mm}
\begin{scriptsize}
	\begin{table}[!htb]
		$$MSPE=\sum_{i=1}^{B}\left( y_{n+h}-\hat{y}_{n+h} \right)^2$$
		\centering
		\begin{tabular}{lll|ccc|ccc|ccc|ccc}
			\hline
			% &  $\phi$&&  & 0 &  &  & 0.3 &  &  & 0.6 &  &  & 0.9 & \\  \hline
			&  &$\phi$&  & 0 &  &  & 0.3 &  &  & 0.6 &  &  & 0.9 & \\  \hline
			$snr$ & Model& $h$ & 1 & 5 & 10 & 1 & 5 & 10 & 1 & 5 & 10 & 1 & 5 & 10 \\ 
			\hline  
			0.05 & LM.PC  && 0.14 & 0.14 & 0.14 & 0.13 & 0.14 & 0.12 & 0.10 & 0.09 & 0.10 & 0.03 & 0.03 & 0.03 \\ 
			0.05 & GLS.PC && 0.14 & 0.14 & 0.14 & 0.12 & 0.14 & 0.12 & 0.07 & 0.09 & 0.10 & 0.01 & 0.02 & 0.03 \\ \hline
			0.05 & LM.BSP  && 0.14 & 0.14 & 0.14 & 0.12 & 0.14 & 0.12 & 0.10 & 0.09 & 0.10 & 0.03 & 0.02 & 0.03 \\ 
			0.05 & GLS.BSP && 0.14 & 0.14 & 0.14 & 0.11 & 0.14 & 0.12 & 0.06 & 0.09 & 0.09 & 0.00 & 0.02 & 0.02 \\  \hline  \hline
			0.10 & LM.PC &&0.34 & 0.33 & 0.28 & 0.28 & 0.28 & 0.26 & 0.21 & 0.20 & 0.21 & 0.06 & 0.06 & 0.07 \\ 
			0.10 & GLS.PC &&  0.34 & 0.33 & 0.28 & 0.25 & 0.28 & 0.26 & 0.14 & 0.19 & 0.21 & 0.02 & 0.05 & 0.06 \\ \hline
			0.10 & LM.BSP &&0.34 & 0.34 & 0.29 & 0.28 & 0.28 & 0.27 & 0.21 & 0.20 & 0.22 & 0.06 & 0.06 & 0.07\\ 
			0.10 & GLS.BSP &&  0.35 & 0.34 & 0.29 & 0.25 & 0.28 & 0.26 & 0.14 & 0.19 & 0.21 & 0.01 & 0.04 & 0.06 \\
			\hline   \hline
			0.20 & LM.PC&&0.61 & 0.76 & 0.73 & 0.64 & 0.59 & 0.58 & 0.45 & 0.42 & 0.48 & 0.14 & 0.14 & 0.14 \\ 
			0.20 & GLS.PC& & 0.60 & 0.76 & 0.73 & 0.58 & 0.58 & 0.58 & 0.30 & 0.41 & 0.47 & 0.04 & 0.10 & 0.12 \\  \hline 
			0.20 & LM.BSP&&0.64 & 0.77 & 0.74 & 0.64 & 0.63 & 0.59 & 0.45 & 0.44 & 0.47 & 0.14 & 0.14 & 0.14 \\ 
			0.20 & GLS.BSP& & 0.64 & 0.78 & 0.74 & 0.59 & 0.63 & 0.58 & 0.29 & 0.40 & 0.46 & 0.03 & 0.09 & 0.11 \\ 
			\hline   \hline
		\end{tabular}
		\caption{Mean squared prediction error for different lags. Model (b)}
		\label{tab:MSE2}
	\end{table}
\end{scriptsize}

\end{document}